\documentclass[epj,nopacs]{svjour}
\usepackage{epsfig}
\usepackage{subfigure}
\usepackage{amssymb}
\usepackage{amsmath}
\usepackage{latexsym}
\begin{document}
\hugehead
\title{Multiplicity of Charged and Neutral Pions in 
Deep-Inelastic Scattering of 27.5~GeV Positrons on Hydrogen}

\author{ 
The HERMES Collaboration \medskip \\ 
A.~Airapetian$^{31}$,
N.~Akopov$^{31}$,
Z.~Akopov$^{31}$,
M.~Amarian$^{23,26,31}$,
J.~Arrington$^{2}$,
E.C.~Aschenauer$^{7,13,23}$,
H.~Avakian$^{11}$,
R.~Avakian$^{31}$,
A.~Avetissian$^{31}$,
E.~Avetissian$^{31}$
% ...
} 

\institute{ 
$^1$Department of Physics, University of Alberta, Edmonton, Alberta T6G 2J1, Canada\\
$^2$Physics Division, Argonne National Laboratory, Argonne, Illinois 60439-4843, USA\\
$^3$Istituto Nazionale di Fisica Nucleare, Sezione di Bari, 70124 Bari, Italy\\
$^4$W.K. Kellogg Radiation Laboratory, California Institute of Technology, Pasadena, California 91125, USA\\
$^5$Nuclear Physics Laboratory, University of Colorado, Boulder, Colorado 80309-0446, USA\\
$^6$DESY, Deutsches Elektronen Synchrotron, 22603 Hamburg, Germany\\
$^7$DESY Zeuthen, 15738 Zeuthen, Germany\\
$^8$Joint Institute for Nuclear Research, 141980 Dubna, Russia\\
$^9$Physikalisches Institut, Universit\"at Erlangen-N\"urnberg, 91058 Erlangen, Germany\\
$^{10}$Istituto Nazionale di Fisica Nucleare, Sezione di Ferrara and Dipartimento di Fisica, Universit\`a di Ferrara, 44100 Ferrara, Italy\\
$^{11}$Istituto Nazionale di Fisica Nucleare, Laboratori Nazionali di Frascati, 00044 Frascati, Italy\\
$^{12}$Fakult\"at f\"ur Physik, Universit\"at Freiburg, 79104 Freiburg, Germany\\
$^{13}$Department of Subatomic and Radiation Physics, University of Gent, 9000 Gent, Belgium\\
$^{14}$Max-Planck-Institut f\"ur Kernphysik, 69029 Heidelberg, Germany\\
$^{15}$Department of Physics, University of Illinois, Urbana, Illinois 61801, USA\\
$^{16}$Physics Department, University of Liverpool, Liverpool L69 7ZE, United Kingdom\\
$^{17}$Department of Physics, University of Wisconsin-Madison, Madison, Wisconsin 53706, USA\\
$^{18}$Laboratory for Nuclear Science, Massachusetts Institute of Technology, Cambridge, Massachusetts 02139, USA\\
$^{19}$Randall Laboratory of Physics, University of Michigan, Ann Arbor, Michigan 48109-1120, USA \\
$^{20}$Lebedev Physical Institute, 117924 Moscow, Russia\\
$^{21}$Sektion Physik, Universit\"at M\"unchen, 85748 Garching, Germany\\
$^{22}$Department of Physics, New Mexico State University, Las Cruces, New Mexico 88003, USA\\
$^{23}$Nationaal Instituut voor Kernfysica en Hoge-Energiefysica (NIKHEF), 1009 DB Amsterdam, The Netherlands\\
$^{24}$Petersburg Nuclear Physics Institute, St. Petersburg, Gatchina,
188350 Russia\\
$^{25}$Institut f\"ur Theoretische Physik, Unversit\"at Regensburg,
93040 Regensburg, Germany\\
$^{26}$Istituto Nazionale di Fisica Nucleare, Sezione Roma 1 - Gruppo Sanit\`a and Physics Laboratory, Istituto Superiore di Sanit\`a, 00161 Roma, Italy\\
$^{27}$Department of Physics, Simon Fraser University, Burnaby, British Columbia V5A 1S6, Canada\\
$^{28}$TRIUMF, Vancouver, British Columbia V6T 2A3, Canada\\
$^{29}$Department of Physics, Tokyo Institute of Technology, Tokyo 152, Japan\\
$^{30}$Department of Physics and Astronomy, Vrije Universiteit, 1081 HV Amsterdam, The Netherlands\\
$^{31}$Yerevan Physics Institute, 375036, Yerevan, Armenia
} 

\date{Received: / Revised version:}

\titlerunning{Multiplicity of Charged and Neutral Pions}
\authorrunning{The HERMES Collaboration}

\abstract{Measurements of the individual multiplicities of $\pi^+$,
$\pi^-$ and $\pi^0$ produced in the deep-inelastic scattering of
27.5~GeV positrons on hydrogen are presented.  The average charged
pion multiplicity is the same as for neutral pions, up to $z \approx
0.7$, where $z$ is the fraction of the energy transferred in the
scattering process carried by the pion.  This result (below $z \approx
0.7$) is consistent with isospin invariance.  The total energy
fraction associated with charged and neutral pions is $\rm 0.51 \pm
0.01 (stat.) \pm 0.08$ (syst.) and $\rm 0.26 \pm 0.01 (stat.) \pm 0.04
(syst.)$, respectively.  For fixed $z$, the measured multiplicities
depend on both the negative squared four momentum transfer $Q^2$ and
the Bjorken variable $x$.  The observed dependence on $Q^2$ agrees
qualitatively with the expected behaviour based on NLO-QCD evolution,
while the dependence on $x$ is consistent with that of previous data
after corrections have been made for the expected $Q^2$-dependence.}

\maketitle

\section{Introduction}

\begin{figure} [ht]
 \begin{center}
  \epsfig{file=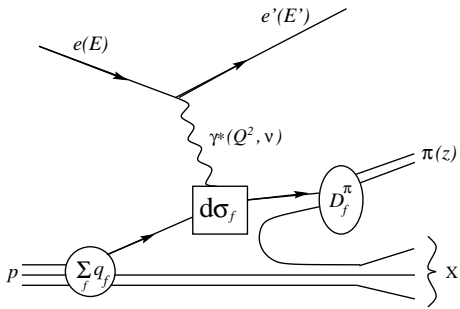,width=6cm}
 \end{center}
 \caption{Semi-inclusive pion electroproduction diagram.}
 \label{fig:fact}
\end{figure}

The semi-inclusive production of pseudoscalar mesons in Deep-Inelastic
Scattering (DIS) is a good tool to test the quark-parton model and
QCD.  A schematic diagram and the relevant variables for the process
are shown in Fig.~\ref{fig:fact}.  In the proton rest frame, the
energy of the exchanged virtual photon $\gamma^*$ is $\nu=E-E'$ ($E$
and $E'$ being the energies of the incident and scattered positrons
respectively), while its squared four-momentum is $-Q^2$.  The
quantity $x= Q^2/2M\nu$, where $M$ is the proton mass, is the fraction
of the light-cone momentum of the nucleon carried by the struck quark.
The parton distribution function $q_f$ describes the momentum
distribution of quarks in the nucleon, while the fragmentation
function $D_f^{\pi}$ is a measure of the probability that a quark of
flavour $f$ fragments into a pion of energy $E_{\pi} = z \nu$.  The
quantity $d\sigma_f$ is the cross-section for the absorption of the
virtual photon by the struck quark.

The quantity of interest in this paper is the pion differential
multiplicity, or the number ($N^\pi$) of pions produced in DIS, 
normalised to the total number ($N_{DIS}$) of inclusive
DIS events ($e + p \rightarrow e' + X$).  In the QCD improved
quark-parton model, it is given by the expression:
\begin{equation}
\frac{1}{N_{DIS}(Q^2)}\frac{dN^{\pi}(z,Q^2)}{dz}=
\frac{\underset{f}{\sum}e^{2}_{f} \, \int_0^1dx \ 
q_{f}(x,Q^2)D^{\pi}_{f}(z,Q^2)}
{\underset{f}{\sum}e^{2}_{f} \, \int_0^1dx \ q_{f}(x,Q^2)},
\label{eq2}
\end{equation}
where the sum is over quarks and antiquarks of flavour $f$, and
$e_{f}$ is the quark charge in units of the elementary charge.
Perturbative QCD calculations in leading \cite{OWE,UEM} and
next-to-leading order \cite{BAI,SAK,CHI,GRE,BKK0,BKK,KKP} suggest a
significant $Q^2$-dependence of the fragmentation process.  These QCD
expectations have been verified by experimental results of hadron
production in e$^+$e$^-$ collisions (see references to experiments in
\cite{BKK0,BKK,KKP}), and in lepton-nucleon scattering at large $\nu$
and $Q^2$ \cite{EMC1,EMC4,BEBC,ZEU,ADL,ADA,ALL,WIT}.  Data with pions
identified in the final state are also available
\cite{EMC2,SLAC,EMC3,ARN,DREWS}.  This paper presents measurements at
lower $Q^2$ of multiplicities for both charged and neutral pions as a
function of the variable $z$ in the range $0.1\le z \le 0.9$.  This
study of the $Q^2$-dependence above 1 GeV$^2$ provides new and precise
information on the scaling violation of the pion fragmentation process
at relatively small values of $W$ ($3.1$ GeV $\le W \le 6.6$ GeV)
\cite{PAS}; $W$ is the invariant mass of the virtual-photon + proton
system.

Evidence for an additional $x$ or $W$ dependence of the multiplicities
has been seen by previous experiments \cite{EMC1,EMC4,BEBC}.  This has
been ignored in the formalism leading to Eq.~\ref{eq2}.  The
multiplicities measured at HERMES were also studied as a function of
$x$ to determine whether they show a similar behaviour.  Also, charged
and neutral pion multiplicities are compared as a test of isospin
invariance.

\section{Experiment}

The measurements described here were performed with the HERMES
spectrometer \cite{SP} using the 27.5 GeV positron beam stored in the
HERA ring at DESY.  The spectrometer consists of two identical halves
above and below the positron and proton beam pipes.  The beam with a
typical current in the range between 10 and 35 mA was incident on a
hydrogen internal gas target \cite{TA}.  The data used in this
analysis were collected during the 1996 and 1997 HERA beam periods.
The target was operated in both unpolarised and longitudinally
polarised configurations, with typical areal densities of
8$\times$10$^{14}$ and 7$\times$10$^{13}$ atoms/cm$^2$, respectively.
The data from the polarised target were analysed by averaging over the
two spin orientations, producing results that are consistent with
those from the unpolarised target.

The scattered positrons and any resulting hadrons were detected
simultaneously by the HERMES spectrometer.  The geometrical acceptance
of the spectrometer is $\pm$(40 - 140)~mrad in the vertical direction
and $\pm170$ mrad in the horizontal direction.

The identification of the scattered positron was accomplished using a
gas threshold \v{C}erenkov counter, a transition radiation detector, a
scintillator hodoscope preceded by two radiation lengths of lead
(preshower counter), and an electromagnetic calorimeter.  This system
provided positron identification with an average efficiency of 98\%
and a hadron contamination of less than 1\%.  Events were selected by
imposing the kinematic restrictions $Q^2\ge1$~GeV$^2$ and $y\le0.85$,
where $y = \nu/ E$ is the virtual photon fractional energy.

The \v{C}erenkov counter was filled with a mixture of 70\% nitrogen
and 30\% perfluorobutane ($\rm C_4 F_{10}$), providing a momentum
threshold of 3.8, 13.6 and 25.8~GeV for pions, kaons, and protons
respectively.  The momentum was restricted to the range between
4.5~GeV and 13~GeV in the data analysis to eliminate possible kaon
or proton contamination of the charged pion sample.  The contribution
from the inefficiency of the \v{C}erenkov detector to the systematic
uncertainty on the charged pion multiplicities was evaluated to be
3.2\% \cite{GEI}.

\begin{figure} [ht]
 \begin{center}
  \epsfig{file=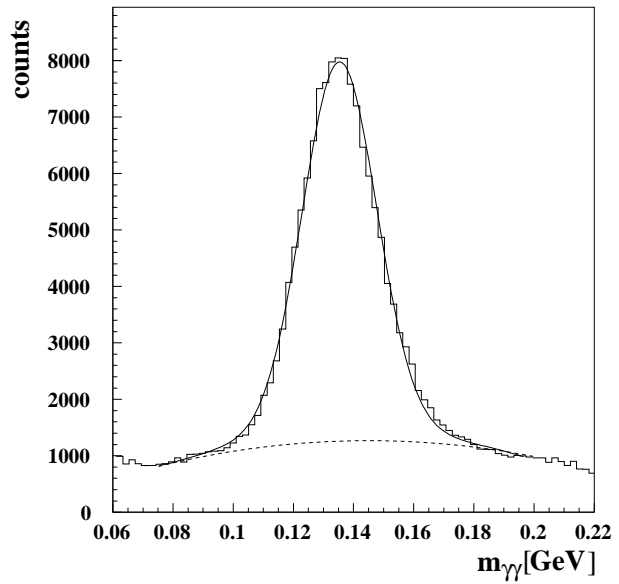,width=8cm}
 \end{center}
 \caption{Two-photon invariant mass spectrum. The solid line is a fit with
a Gaussian plus a polynomial.  The dashed line represents the background
only.}
 \label{fig:pi0}
\end{figure}

The electromagnetic calorimeter is composed of 840 F101 lead glass
blocks \cite{calo}, and provided neutral pion identification by the
detection of the two neutral clusters from two-photon $\pi^0$ decay.
Each of the two clusters was required to have an energy
$E_{\gamma}\ge1.4$ GeV.  The measurement of both the energies
($E_{\gamma1}$ and $E_{\gamma2}$) and the relative angle
($\Theta_{\gamma\gamma}$) of the two photons allowed the
reconstruction of the invariant mass $m_{\gamma\gamma}=\sqrt{4
E_{\gamma1}E_{\gamma2} \sin^2 (\Theta_{\gamma\gamma}/2)}$.  A typical
measured spectrum of $m_{\gamma\gamma}$ is shown in
Fig.~\ref{fig:pi0}.  The distribution is centered at
$m_{\gamma\gamma}=0.1348\pm0.0008$~GeV.  The good $\pi^0$ mass
resolution of about 0.012~GeV allowed a safe background subtraction.
The background was evaluated in each kinematic bin by fitting the
invariant mass spectrum with a Gaussian plus a polynomial that
reproduces well the shape of the background due to uncorrelated
photons.  The relevant contribution to the systematic uncertainty is
less than 2\%. It was evaluated by repeating the fitting procedure for
different ranges and with polynomials of different order \cite{PAS}.
The number of $\pi^0$ detected was obtained by integrating the peak
corrected for background, over the range $\pm 2.5 \sigma$ around the
centroid of the Gaussian.

In order to exclude effects from nucleon resonances as well as
kinematic regions with inadequate geometrical acceptance, the
additional requirement $W^2\ge10$~GeV$^2$ was imposed for this
analysis.  This also helps to select hadrons originating from
fragmentation of the struck quark, by excluding the kinematic region
in which current and target fragments are not well separated.
Fragments from the target remnant are already strongly reduced
due to the forward angle acceptance of the HERMES spectrometer.

After background subtraction and data quality plus kinematic cuts, a
total of 4.2$\times$10$^5$ (1.3$\times$10$^5$) semi-inclusive charged (neutral)
pions were considered in the analysis.

\begin{figure}[ht]
 \begin{center}
  \epsfig{file=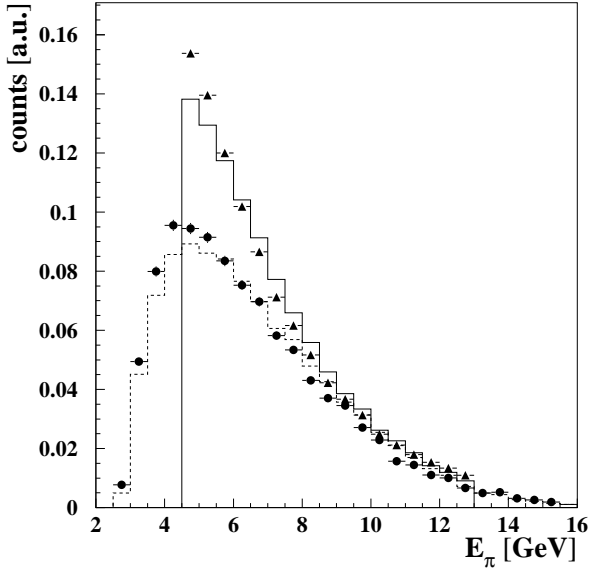,width=8cm}
  \caption{Comparison between the measured and simulated pion energy
spectra.  Filled circles (triangles) are the neutral (charged) pion
measurements.  The histograms are the simulated spectra for neutral
(dashed line) and charged (full line) pions. All spectra have been
normalized to unit area.}
  \label{fig:mc}
 \end{center}
\end{figure}

The HERMES Monte Carlo program (HMC) was used to evaluate the
detection probability for pions produced in DIS events.  HMC is based
on the LEPTO event generator \cite{LEP}, the LUND fragmentation model
\cite{JET}, and on the GEANT3 code for the simulation of the detector
response \cite{GEANT}.  The LUND parameters were adjusted to fit
various kinematic distributions from HERMES semi-inclusive data
\cite{TAL}.  As an example, Fig.~\ref{fig:mc} shows comparisons
between the measured energy spectra of charged and neutral pions and
the relevant HMC simulations.  The agreement in the shape of the
spectra is reasonable. The energy range of the charged pions is
smaller than for neutral pions due to the stronger kinematic
restriction imposed on the data to ensure good charged pion
identification.

\begin{figure}[ht]
 \begin{center}
    \epsfig{file=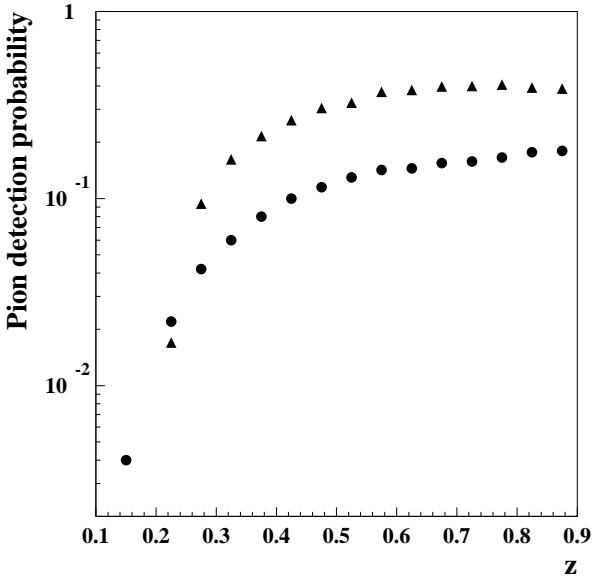,width=8cm}
   \caption{Detection probability as a function of $z$
for charged (triangles) and neutral (circles) pions when the DIS
positron is detected.}
 \label{fig:eff}
  \end{center}
\end{figure}

The detection probability of the pions produced in detected DIS events
is shown as a function of $z$ in Fig.~\ref{fig:eff}.  This quantity
was computed using HMC to account for effects from pion losses due to
the finite angular acceptance of the spectrometer, from detector
inefficiencies, and from fiducial cuts imposed in the data analysis. A
probability up to $\sim$40\% is obtained for charged pions in the
intermediate and high $z$-region.  At low $z$ the detection
probability for charged pions is further reduced due to the
restriction on the pion momentum range.  The probability for neutral
pions is lower due to the need to detect both of the decay photons.
Since the angle of these two photons is correlated with the $\pi^0$
energy, the detection probability strongly increases with $z$,
reaching $\sim$18\% in the highest $z$-bin.  The positron detection
probability does not affect the multiplicity defined in Eq.~\ref{eq2}.
Smearing by instrumental resolution of the positron kinematics in the
various $\nu$-bins was included in the Monte Carlo. This effect also
leads to migration of events in $z$.  Corrections due to smearing vary
from $\sim$2\% in the lowest $z$-bin to 30-35\% in the highest
$z$-bin. The systematic uncertainty on the multiplicities due to the
detection probability (including smearing) is 7.5\% (4.5\%) for
neutral (charged) pions.  It was evaluated with HMC by drastically
restricting the geometrical cuts or the allowed ranges of the relevant
kinematic variables. The multiplicities plotted in this paper were
corrected for the effects described above.

Radiative corrections for internal bremsstrahlung processes were
applied. These corrections range from $\sim$3\% in the lowest $z$-bin
to $\sim$15\% in the highest $z$-bin, and result in a systematic
uncertainty on the multiplicities of less than 1\%.  Radiative
corrections for external bremsstrahlung processes in the target were
found to be negligible due to the small thickness of the hydrogen gas
target \cite{AKU}.

It has been determined from Monte Carlo studies that diffractive
production of pions is negligible in the HERMES kinematic range,
and so no correction was made for this contribution.

\section{Results}
\label{results}

\begin{figure}
 \begin{center} 
     \subfigure{\epsfig{figure=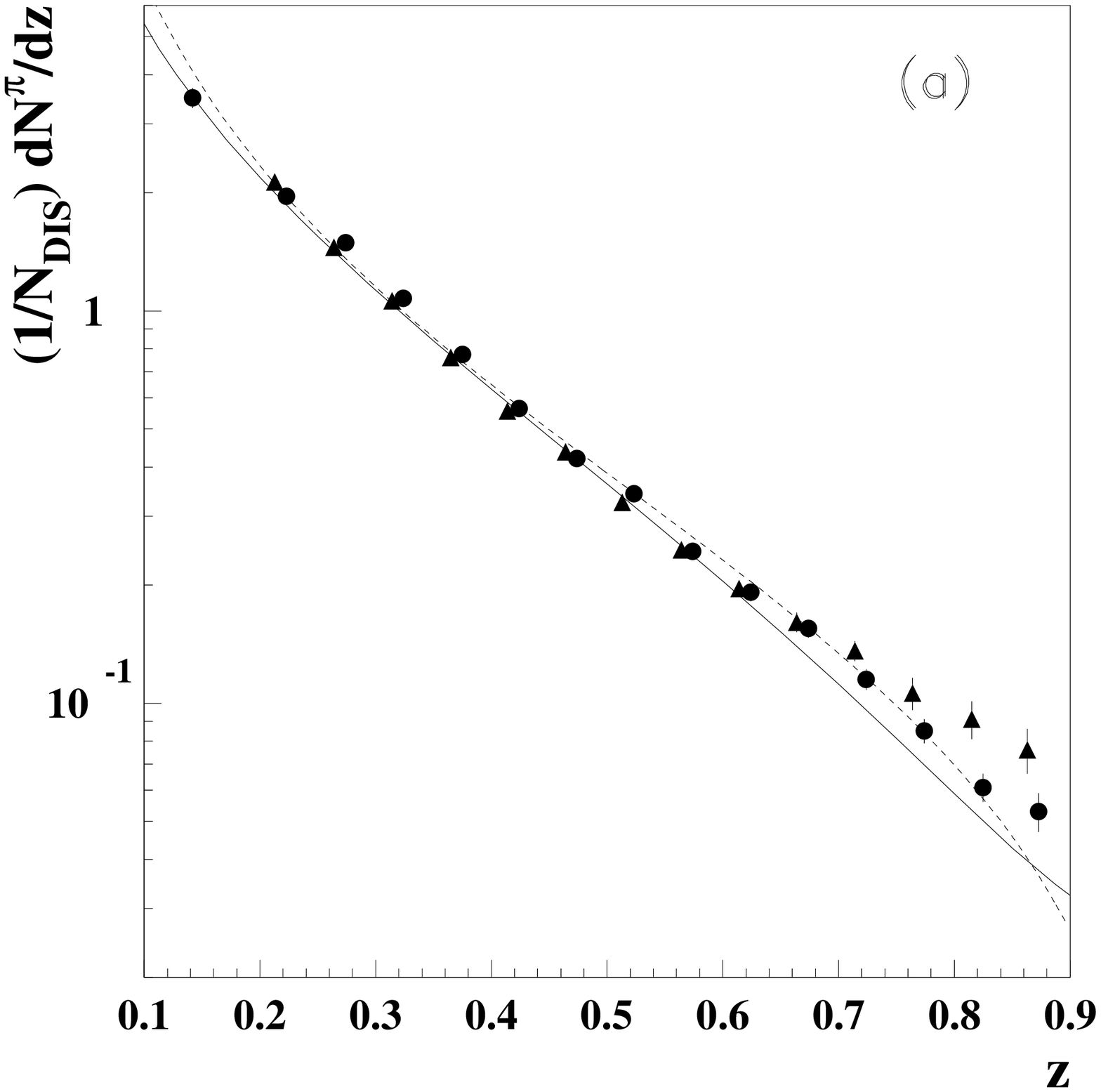,width=8cm}} \\
     \subfigure{\epsfig{figure=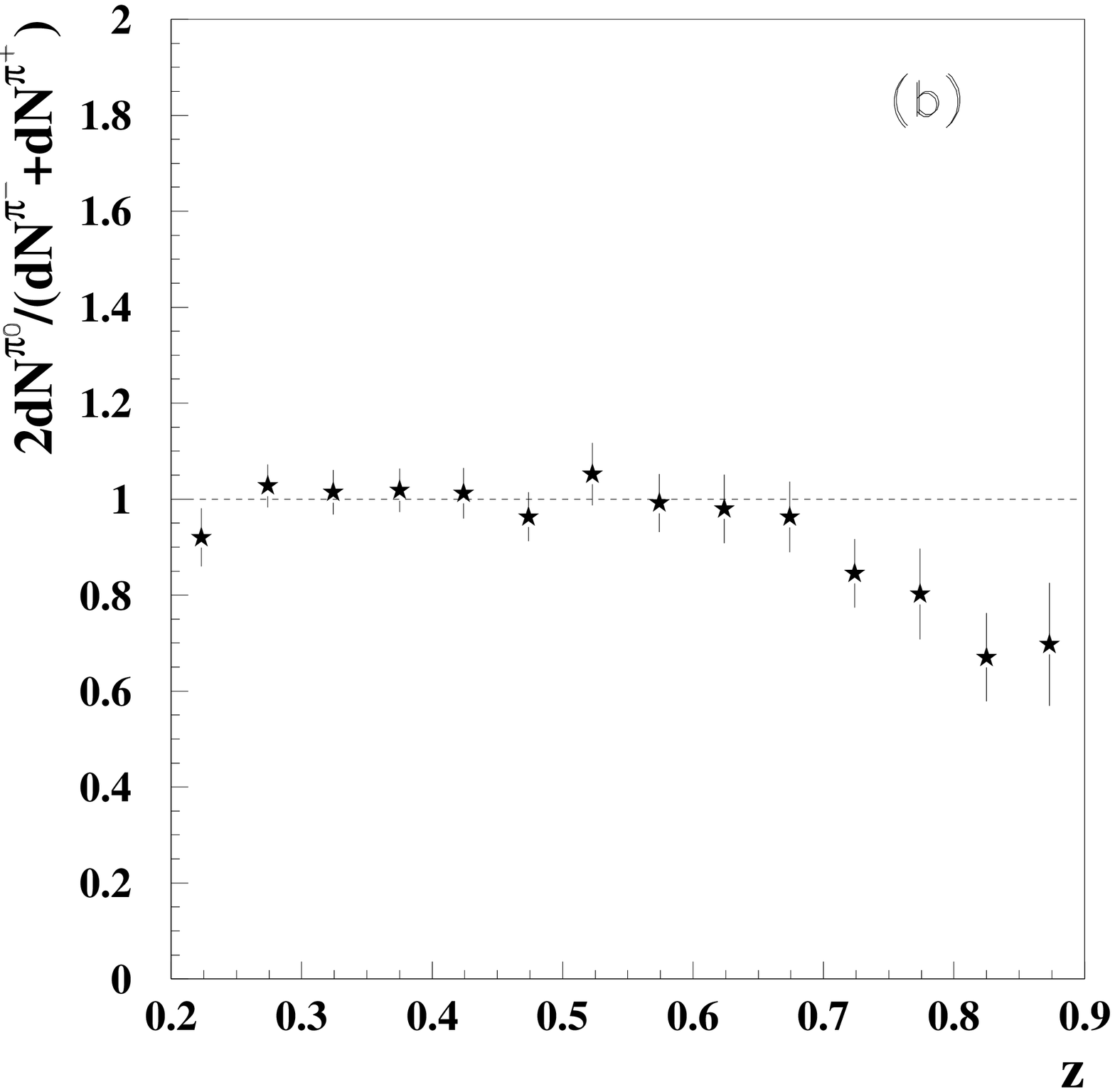,width=8cm}}
\caption{(a) Neutral (circles) and average charged (triangles) pion
multiplicities.  The error bars show the statistical uncertainty only.
The systematic uncertainty for the charged (neutral) pions is 7\%
(9\%).  The solid line is a parameterization using the independent
fragmentation model \cite{FF}.  The dashed line is a fit to the
present neutral pion data using the parameterization given in the
text.  The charged pion data have been shifted slightly in $z$ to make
them visible.  (b) Ratio of neutral to average charged pion
multiplicities.  The systematic uncertainty on the ratio (not included
in the error bar) is 6\%.}
\label{fig:iso}
\end{center}
\end{figure}

Under the assumption of isospin invariance, the quark-parton model
predicts that the multiplicity for neutral pions is equal to the
average of those for positive and negative pions.  Specifically, the
fragmentation function $D_f^{\pi^0}$ is assumed equal to the average
of the two charged pion fragmentation functions $D_f^{\pi^+}$ and
$D_f^{\pi^-}$, because the quark content of the $\pi^0$ is the same as
the average of $\pi^+$ and $\pi^-$.\footnote{$\pi^0=
\frac{1}{\sqrt{2}} (u\bar{u} - d\bar{d})$; $\pi^+= u\bar{d}$; $\pi^-=
d\bar{u}$} The multiplicities $\frac{1}{N_{DIS}}\frac{dN^{\pi^0}}{dz}$
and
$\frac{1}{N_{DIS}}[\frac{dN^{\pi^+}}{dz}+\frac{dN^{\pi^-}}{dz}]/2$,
are plotted as a function of $z$ in Fig.~\ref{fig:iso}(a).  Numerical
values are given in Tab.~\ref{Tab:xsec}, where the individual
multiplicities for $\pi^+$ and $\pi^-$ are also listed, along with the
ranges and average values of the relevant kinematic variables. Both
data sets in Fig.~\ref{fig:iso}(a) show a strong decrease with the $z$
variable that can be parameterized with the expression~\cite{BKK0}
\begin{equation}
\frac{1}{N_{DIS}}\frac{dN^{\pi}}{dz} = N z^{\alpha} (1-z)^{\beta}.
\label{param}
\end{equation}

The dashed curve shown in Fig.~\ref{fig:iso}(a) is a $Q^2$ independent
fit to the $\pi^0$ data using the above expression ($N$ = 0.335,
$\alpha=-1.371$, $\beta=1.167$).  Also shown in Fig.~\ref{fig:iso}(a)
is another $Q^2$-independent parameterization using the independent
fragmentation model.  The coefficients of this parameterization were
obtained in Ref.~\cite{FF} by tuning on old data at $Q^2$-values close
to that of HERMES (see \cite{FF} for references to these data).  This
parameterization reproduces the behaviour of the present data fairly
well, apart from the high $z$-region.  As expected from isospin
invariance, the agreement between the data for neutral and charged
pions is excellent, at least up to $z \sim0.70$ as shown by the ratio
in Fig.~\ref{fig:iso}(b), which is consistent with unity below $z
\sim0.70$.  The verification of isospin invariance agrees with the
conclusion from earlier studies by BEBC, which used neutrino
scattering~\cite{ALL,WIT}, and by EMC~\cite{EMC2} which, however,
compared charged hadrons with neutral pions.  The current
data show that at higher $z$, the multiplicity is larger for charged
pions than for neutral pions. This difference suggests a possible
contribution, via the radiative tail and instrumental resolution, from
exclusive processes (e.g. $\gamma^* + p \rightarrow \pi^+ + \Delta^0$)
where resonances affect each isospin channel differently.  A recent
calculation \cite{FRANK} of hard exclusive electroproduction indeed
predicts an enhancement of charged over neutral pion production of
roughly an order of magnitude.  Higher twist processes could also play
a role at high $z$ \cite{HAROUT}.  The excess of charged over neutral
pions at high $z$ is being investigated further.

Comparison of results obtained with unpolarised and polarised targets
and of those with the top half and the bottom half of the spectrometer
resulted in consistency within 3.5\% and 2.5\% respectively.  It
should be noted that the data sets for charged and neutral pions were
obtained with very different event reconstruction procedures,
detection efficiencies and background conditions.  Hence the
contributions to the systematic uncertainty are quite different for
the two cases.  The overall systematic uncertainty was estimated
to be less than 9\% (7\%) for neutral (charged) pion
electroproduction.  The systematic uncertainty on the ratio plotted in
Fig.~\ref{fig:iso}(b) is 6\%.

\begin{table*}[h]
\begin{center}
\begin{tabular}{|c|c|c|c|c|}\hline
 $z$ & $\frac{1}{N^{DIS}} N^{\pi^{0}}$ & $\frac{1}{N^{DIS}} 
\frac{N^{\pi^+} + N^{\pi^-}}{2}$ & 
$\frac{1}{N^{DIS}} N^{\pi^{+}}$ & $\frac{1}{N^{DIS}} N^{\pi^{-}}$ \\ \hline
0.142 & 3.39$\pm$0.21   &  -  & - & - \\
0.223 & 1.96$\pm$0.08   & 2.13$\pm$0.11 & 2.22$\pm$0.07 & 2.04$\pm$0.08 \\
0.274 & 1.49$\pm$0.05   & 1.45$\pm$0.04 & 1.63$\pm$0.02 & 1.28$\pm$0.02 \\
0.324 & 1.075$\pm$0.038 & 1.06$\pm$0.03 & 1.22$\pm$0.02 & 0.897$\pm$0.023 \\
0.375 & 0.774$\pm$0.028 & 0.760$\pm$0.022 & 0.906$\pm$0.017 & 0.614$\pm$0.014\\
0.424 & 0.563$\pm$0.021 & 0.556$\pm$0.020 & 0.670$\pm$0.015 & 0.442$\pm$0.013\\
0.474 & 0.421$\pm$0.016 & 0.437$\pm$0.016 & 0.532$\pm$0.011 & 0.343$\pm$0.012\\
0.523 & 0.342$\pm$0.014 & 0.325$\pm$0.015 & 0.411$\pm$0.012 & 0.239$\pm$0.009\\
0.574 & 0.244$\pm$0.010 & 0.246$\pm$0.011 & 0.304$\pm$0.009 & 0.188$\pm$0.007\\
0.624 & 0.192$\pm$0.010 & 0.196$\pm$0.010 & 0.250$\pm$0.009 & 0.143$\pm$0.007\\
0.674 & 0.155$\pm$0.008 & 0.161$\pm$0.009 & 0.198$\pm$0.008 & 0.126$\pm$0.007\\
0.724 & 0.115$\pm$0.007 & 0.136$\pm$0.008 & 0.164$\pm$0.007 & 0.108$\pm$0.007\\
0.774 & 0.085$\pm$0.006 & 0.106$\pm$0.010 & 0.137$\pm$0.007 & 0.075$\pm$0.005\\
0.825 & 0.061$\pm$0.005 & 0.091$\pm$0.010 & 0.112$\pm$0.008 & 0.069$\pm$0.007\\
0.873 & 0.053$\pm$0.006 & 0.076$\pm$0.010 & 0.100$\pm$0.009 & 0.054$\pm$0.008\\
\hline
\end{tabular}
\caption{Measured $\pi^0$, ($\pi^+ + \pi^-$)/2, $\pi^+$, and $\pi^-$,
multiplicities for various $z$-bins. The average kinematic variables
for this experiment are: $E_{beam}=$ 27.5~GeV, $\sqrt{s}=$ 7.48~GeV,
$\langle Q^2 \rangle =$ 2.5~GeV$^2$,
$\langle W^2 \rangle =$ 28.6~GeV$^2$,
$\langle \nu \rangle =$ 16.1~GeV,
$\langle x \rangle =$ 0.082, while the ranges are
$Q^2$= [1, $\sim 15$] GeV$^2$, $W^2$= [10, $\sim 44$] GeV$^2$, 
$\nu$= [$\sim 5.4$, 23.4] GeV, and $x$= [0.03, $\sim 0.6$].
The quoted uncertainties are statistical. The systematic uncertainty
for the charged (neutral) pions is 7\% (9\%).}
\label{Tab:xsec}
\end{center}
\end{table*}

The fraction of the energy $\nu$ of the virtual photon transfered to pions
\begin{equation}
\int_{0}^{1} z \frac{1}{N_{DIS}}\frac{dN^{\pi}}{dz}dz
\label{eq3}
\end{equation}
is $\rm 0.26 \pm 0.01 (stat.) \pm 0.04 (syst.)$ and $\rm 0.51 \pm 0.01
(stat.) \pm 0.08 (syst.)$, for the neutral and charged cases,
respectively.  The corresponding number for neutral pions measured by
EMC \cite{EMC2} is $0.27 \pm 0.02 \pm 0.05$.  The current results
indicate that the fraction of total energy carried by hadrons heavier
than pions (mainly $K^+$, $K^-$, $p$, $\bar{p}$) is only $\sim23\%$.
The integrals of Eq.~\ref{eq3} were evaluated by adding estimates of
the contributions from the unmeasured $z$-regions to those from the
measured $z$-region, the latter contributions being $\rm 0.19 \pm 0.01
(stat.)  \pm 0.017 (syst.)$ for neutral and $\rm 0.28 \pm 0.01 (stat.)
\pm 0.02 (syst.)$ for charged pions.  This extrapolation was based on
the $z$-dependence given by the fit of ref.~\cite{FF} (solid line in
Fig.~\ref{fig:iso}(a)).  The total systematic uncertainty includes an
estimate of the error in the extrapolation evaluated by comparing
these results with those obtained using the fit of Eq.~\ref{param}.
This contribution, 14\% (13\%) for charged (neutral) pions, dominates
the overall systematic uncertainty.  It is assumed in estimating this
error that the models used are a reasonable representation of the
data, even at small $z$.  The possibility of a radically different
behaviour of the multiplicities at small $z$ is not accounted for in
the systematic uncertainty.

\begin{figure}
 \begin{center}
    \subfigure{\epsfig{figure=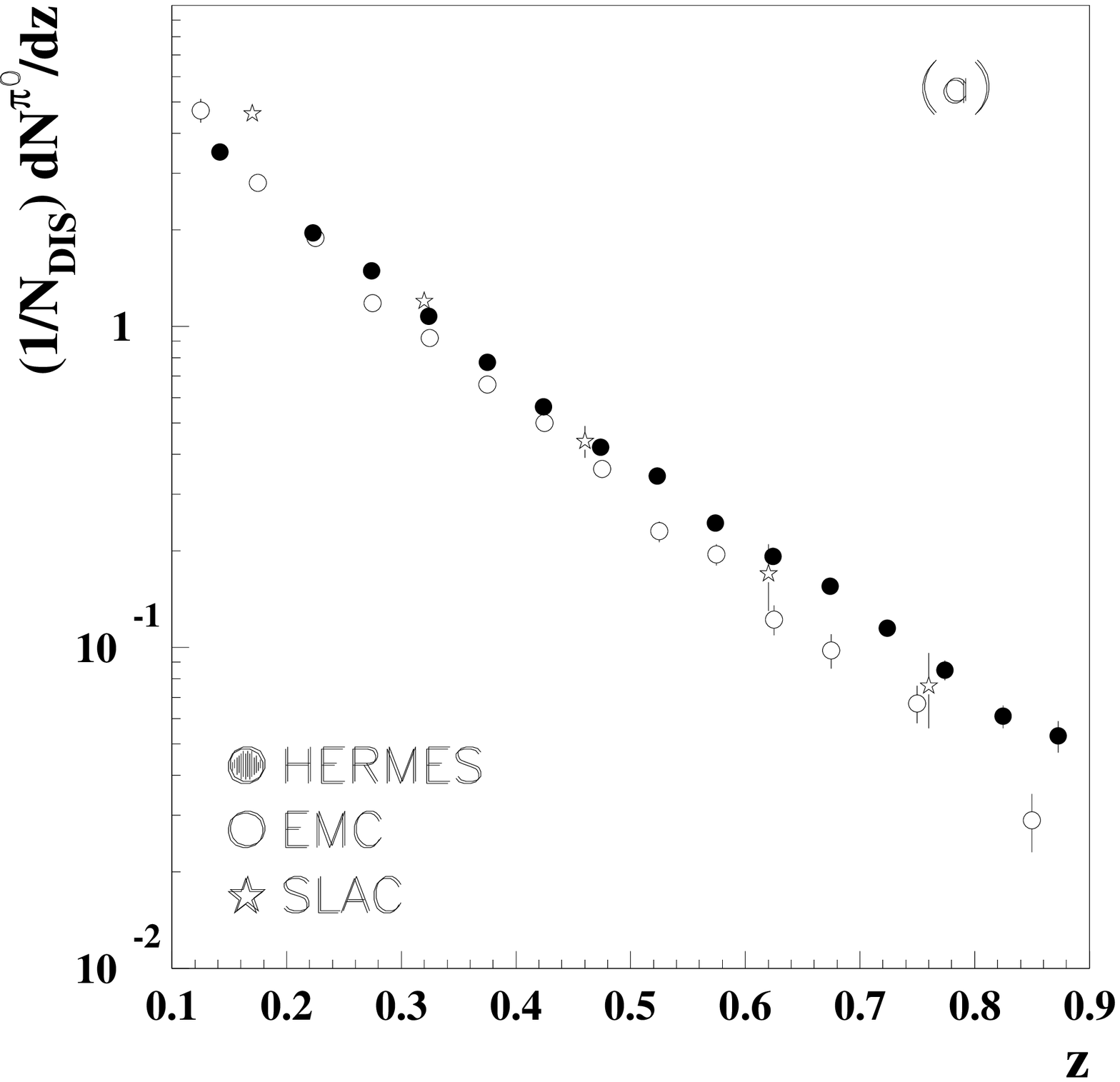,width=8cm}} \\
    \subfigure{\epsfig{figure=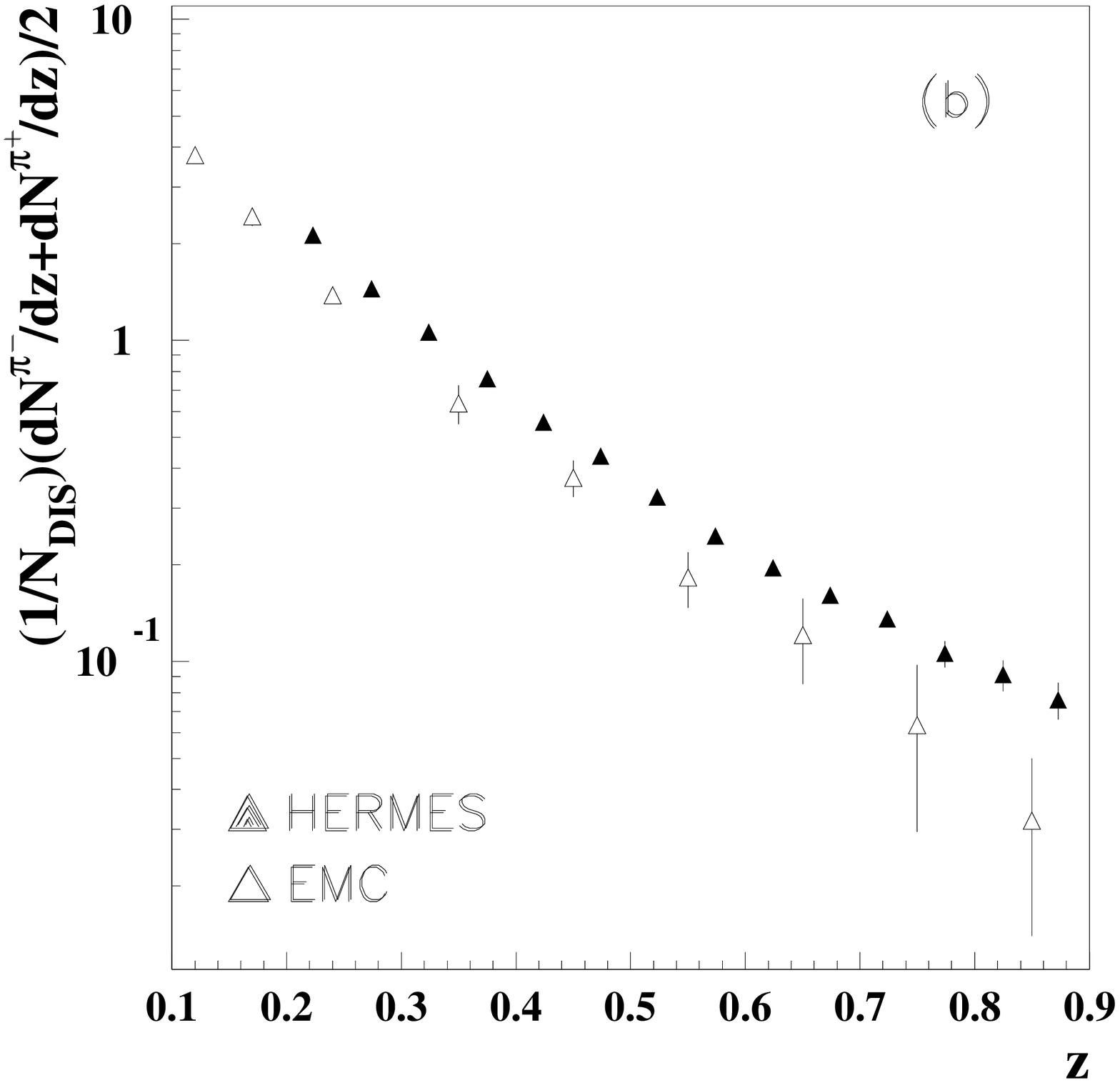,width=8cm}}
\caption{(a) $\pi^{0}$ multiplicity from HERMES, EMC
\protect\cite{EMC2} and SLAC \protect\cite{SLAC}.  (b) Average charged
pion multiplicity from HERMES compared to EMC fragmentation functions
\protect\cite{EMC3}. Only the statistical uncertainties are shown. The
systematic uncertainties for neutral (charged) pions are $9\%$ ($7\%$)
for HERMES, $\le 15\%$ for SLAC and $\le 13\%$ ($\le 10\%$) for EMC.}
 \label{fig:cfr}
  \end{center}
\end{figure}

\begin{figure}
 \begin{center}
    \subfigure{\epsfig{figure=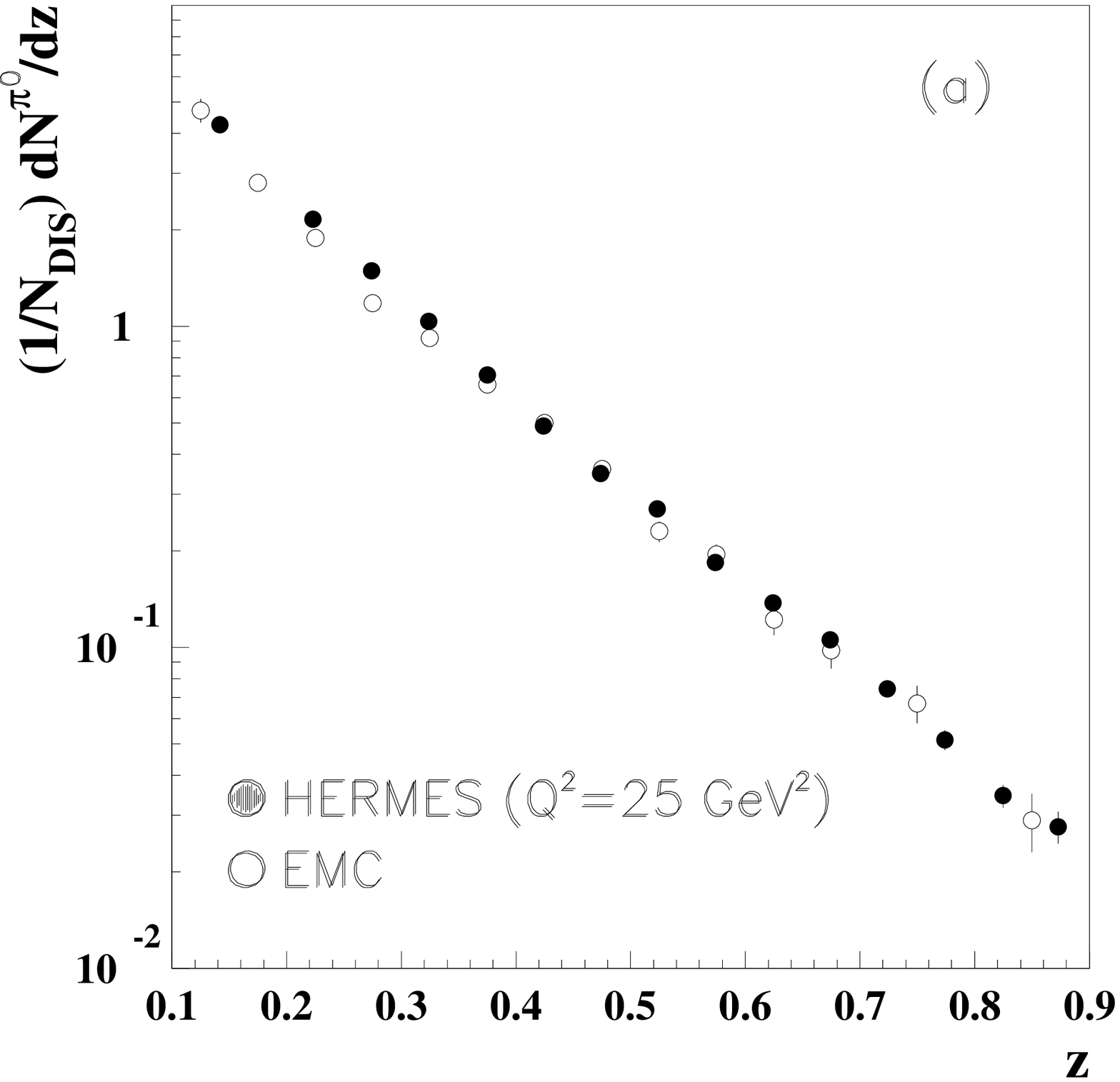,width=8cm}} \\
    \subfigure{\epsfig{figure=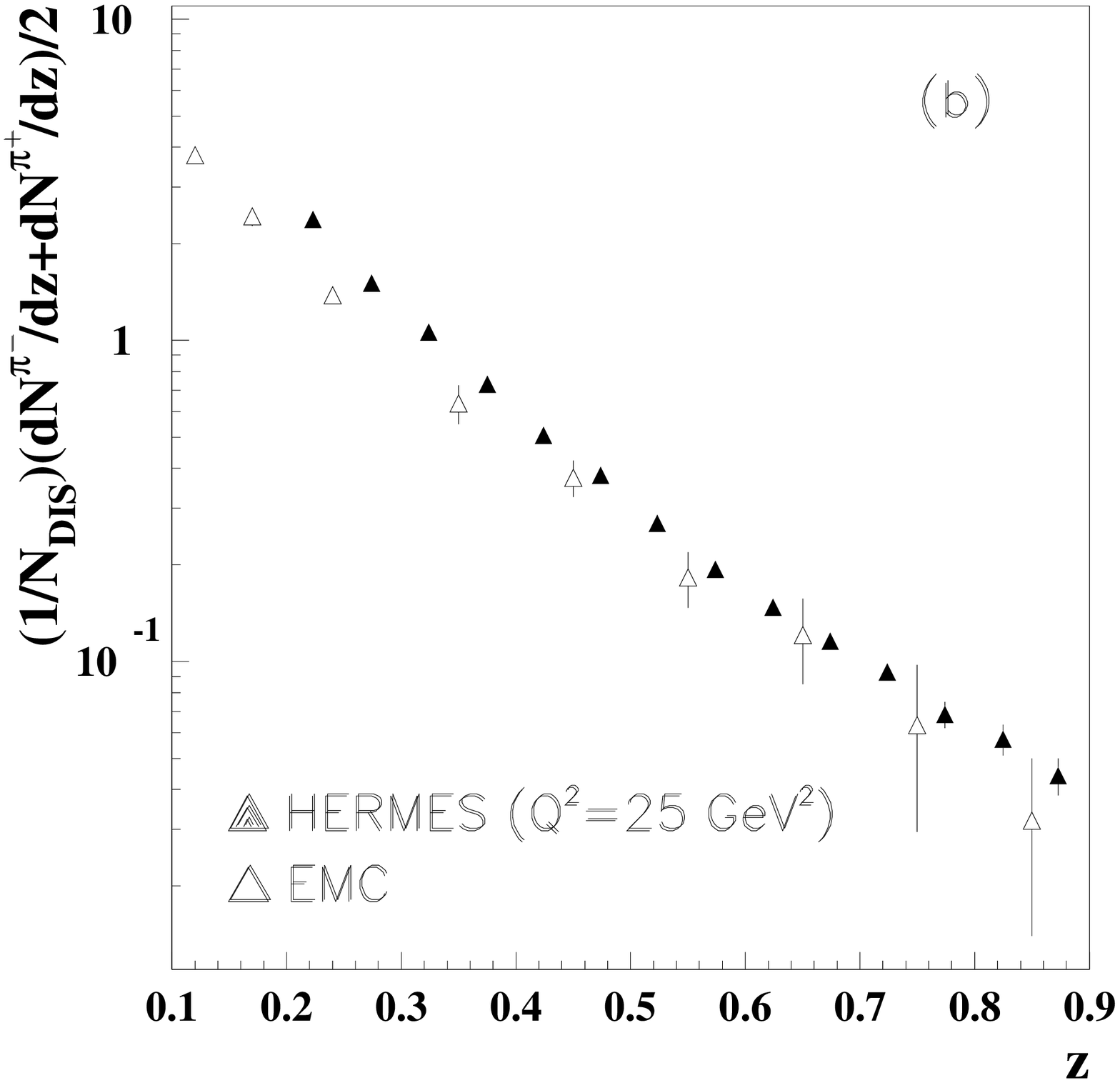,width=8cm}}
 \caption{Multiplicities for (a) neutral pions, and (b) the average of
charged pions (note the EMC data for charged pions are for
fragmentation functions).  The HERMES results have been evolved to
$Q^{2}=$25 GeV$^2$ using a NLO QCD model~\protect\cite{KKP}.  Only
statistical uncertainties are shown. The systematic uncertainties for
neutral (charged) pions are $9\%$ ($7\%$) for HERMES and $\le 13\%$
($\le 10\%$) for EMC.}
 \label{fig:cfr2}
 \end{center}
\end{figure}

In Fig.~\ref{fig:cfr}(a) the HERMES results for the multiplicity of
neutral pions as a function of $z$ are compared with previous results
from EMC \cite{EMC2} and from SLAC \cite{SLAC}.  Fig.~\ref{fig:cfr}(b)
compares HERMES results for the average charged pion multiplicity with
fragmentation function data from EMC \cite{EMC3}.  The HERMES
results for both neutral and charged pions are systematically higher
than those from EMC.  This difference can be explained by the
different $Q^2$ range covered by the two experiments: $\langle Q^2
\rangle =$ 2.5 GeV$^2$ for HERMES and $\langle Q^2 \rangle =$ 25
GeV$^2$ for EMC.  The $Q^2$ range for the SLAC data is 1.8-8.5
GeV$^2$.  Similar $Q^2$-dependent behaviour has been seen for
hadrons in ref.~\cite{ADA}. 
In Figs.~\ref{fig:cfr2}(a) and \ref{fig:cfr2}(b) the HERMES
data have been evolved to the mean $Q^2$ of the EMC data using an NLO
model for the evolution of the fragmentation functions \cite{KKP}.
The agreement between the evolved HERMES data and the EMC data is much
improved, especially for $\pi^0$, demonstrating the need for QCD
corrections.  The comparison of multiplicities and fragmentation
functions, and perhaps to a better extent the application to
multiplicities of the model for the evolution of fragmentation
functions, can be justified if isospin symmetry is assumed ($D_u^{\pi}
= D_d^{\pi}$; where $\pi$ represents the sum of positive and negative
pions) and the strange quark contribution is neglected.  In this case,
the multiplicities are equivalent to the fragmentation functions.  The
effect of ignoring the strange quark can be inferred from a study in
Ref.~\cite{GEI}, where the effect of neglecting all sea quarks was
estimated to be less than 10\% (between 20\% and 40\%) for favoured
(disfavoured) fragmentation functions.  The dominance of $u$-quarks
and the fact that the strange quark content is expected to be less
than the light sea leads to the conclusion that the effect of the
strange quarks should be well below 10\%.  On the other hand, the fact
that the HERMES multiplicities are larger than the EMC fragmentation
functions at low $z$, even after $Q^2$ corrections (see
Fig.~\ref{fig:cfr2}(b)) could indicate a failure of the above
approximation, which is more likely at low $z$.

\begin{figure}
 \begin{center} \epsfig{figure=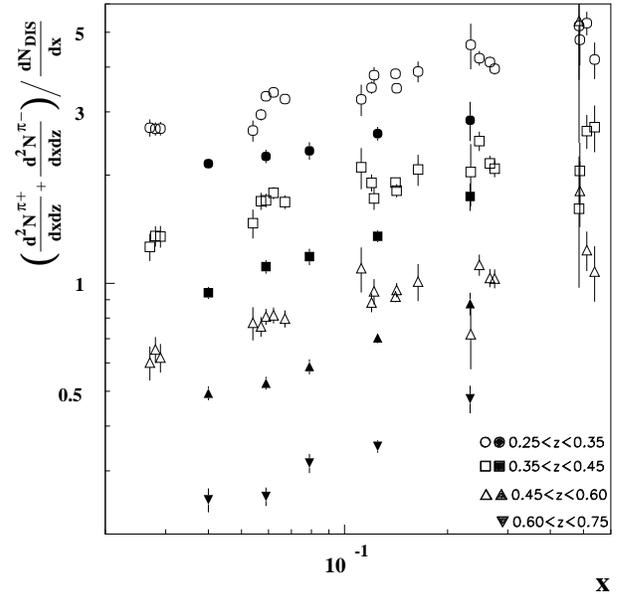,width=8cm}
   \caption{Charged pion multiplicities $(\frac{d^2 N^{\pi+}}{dx\,dz}
+ \frac{d^2 N^{\pi-}}{dx\,dz}) / \frac{d N_{DIS}}{dx}$ from HERMES
(filled symbols) as a function of $x$ in four different $z$-bins,
compared to charged hadron multiplicities from EMC (open symbols). All
data have been evolved to 2.5~GeV$^2$ and are plotted at the measured
$x$-value.}
\label{fig:xdep} \end{center}
\end{figure}

The charged pion multiplicities for HERMES are plotted in
Fig.~\ref{fig:xdep} as a function of $x$ in four bins of $z$ together
with data on charged hadron multiplicities from EMC\footnote{The EMC
pion data are not available in the required kinematic binning.}
\cite{EMC4}.  All data have been evolved to $Q^2$=~2.5~GeV$^2$, the
average $Q^2$ of HERMES, using the model described in the previous
paragraph.  The data are plotted at the measured $x$-value.  The
difference in the absolute value of the multiplicities is simply
related to the fact that the HERMES data are for pions only while the
EMC data are for hadrons.  A significant $x$-dependence is seen, which
gets stronger as $z$ increases.  The mean $Q^2$ per bin for HERMES
varies only between 2.1 and 2.6~GeV$^2$.  The observed $x$-dependence
is therefore not generated by the $Q^2$ correction.  It is striking
that the slopes in the data from both experiments are consistent even
though they were measured at very different kinematics.  The
$x$-dependence is not expected to affect the $Q^2$ evolution of the
data integrated over the measured $x$ range.  This is bourne out by
the good agreement after NLO-QCD corrections of the two data sets for
$\pi^0$ multiplicities in Fig.~\ref{fig:cfr2}(a).  As stated earlier,
the poorer agreement in Fig.~\ref{fig:cfr2}(b) for charged pions could
be due to the fact that multiplicities are compared to fragmentation
functions in this case.

\begin{figure} [ht]
 \begin{center} \epsfig{figure=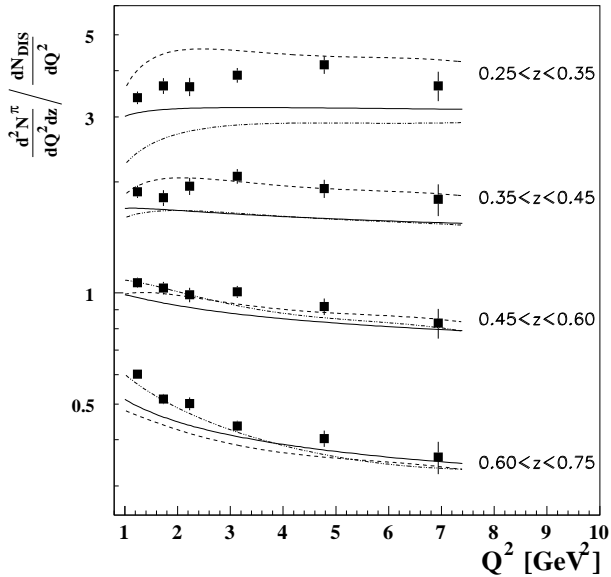,width=8cm}
   \caption{Total (neutral plus charged) pion multiplicity 
   $\frac{d^2 N^{\pi}}{dQ^2\,dz} / \frac{d N_{DIS}}{dQ^2}$
   as a function of $Q^{2}$ for various $z$-bins.  The systematic
   uncertainty on the data is 8.5\%. The three curves shown are
   NLO QCD calculations of fragmentation functions found in
   refs.~\protect\cite{BKK0} (dashed line), \protect\cite{BKK}
   (dotted line), and \protect\cite{KKP} (solid line).}
  \label{fig:vf1} \end{center}
\end{figure}

In order to better illustrate scaling violations in the fragmentation
process, the $Q^2$-dependence of the data at fixed $z$ was studied.
For this analysis, only the data in the range $0.03\le x \le 0.3$ were
considered, to allow for a useful range of $Q^2$, which is quite
limited by the HERMES acceptance for certain values of $x$.  The total
(neutral plus charged) pion multiplicity is plotted for 4 different
$z$-bins in Fig.~\ref{fig:vf1}.  The data show a clear
$Q^2$-dependence, especially in the high $z$-bins.  The variation of
the data with $Q^2$ (i.e. the slope of the curves) is mostly in
agreement with the $Q^2$-evolution predicted by the NLO QCD models
\cite{BKK0,BKK,KKP}.  It is worth mentioning that the above
calculations are based on three analyses of fragmentation functions
extracted from $e^+e^-$ data taken at center of mass energies of 29
GeV \cite{BKK0} and of 29 and 91 GeV \cite{BKK,KKP}, i.e. at much
higher energies than that of the data shown in Fig.~\ref{fig:vf1}.
The sophistication of the parameterization increases for each
successive version of the model, with the solid curve representing the
most recent analysis.  The absolute values of the present measurements
are for the most part in reasonable agreement with the QCD
predictions, considering that each theoretical calculation has an
uncertainty of up to 15\%, the systematic uncertainty on the data is
8.5\%, and that data on multiplicities are compared to a
parameterization of fragmentation functions.

\section{Conclusion}

Charged and neutral pion multiplicities in semi-inclusive
deep-inelastic scattering at 27.5~GeV have been measured by the HERMES
collaboration. The multiplicities are consistent with isospin
invariance below $z\sim 0.70$, and show that about 3/4 of the energy
transferred in the scattering is carried by pions. These measurements
provide data at lower $Q^2$ with improved statistical and systematic
accuracies compared to earlier measurements.  The agreement of the
current results with previous data is improved, especially for
$\pi^0$, when an NLO $Q^2$-evolution of the fragmentation process is
taken into account.  The $Q^2$-behaviour of the present data resembles
that of the NLO QCD calculations.  An observed $x$-dependence of the
multiplicities is similar to previous results on hadron production
from EMC when the data are evolved to the same $Q^2$.

\section{Acknowledgments}

We thank M.~Greco, G.~Kramer, and B.~P\"otter for useful discussions
on fragmentation functions.  We gratefully acknowledge the DESY
management for its support and the staffs at DESY and the
collaborating institutions for their significant effort.  This work
was supported by the Fund for Scientific Research-Flanders (FWO) of
Belgium; the Natural Sciences and Engineering Research Council of
Canada; the INTAS, HCM and TMR contributions from the European
Community; the German Bundesministerium f\"{u}r Bildung, Wissenschaft,
Forschung und Technologie (BMBF), the Deutscher Akademischer
Austauschdienst (DAAD); the Italian Istituto Nazionale di Fisica
Nucleare (INFN); Monbusho, JSPS, and Toray Science Foundation of
Japan; the Dutch Stichting voor Fundamenteel Onderzoek der Materie
(FOM); the UK Particle Physics and Astronomy Research Council; and the
US Department of Energy and National Science Foundation.

\end{document}